\documentclass[preprint,times]{aastex63} %, linenumbers
\usepackage{CJK}
\usepackage{amsmath}
\usepackage{mathtools}
\usepackage[caption=false]{subfig}

\shorttitle{DCF compilation}
\shortauthors{Liu et al.}

\begin{document}
\begin{CJK*}{UTF8}{gbsn}
\title{Magnetic fields in star formation: a complete compilation of all the DCF estimations}

\correspondingauthor{Junhao Liu}
\email{liujunhao42@outlook.com; j.liu@eaobservatory.org}

\author[0000-0002-4774-2998]{Junhao Liu (刘峻豪)}
\affil{East Asian Observatory, 660 N. A`oh\={o}k\={u} Place, University Park, Hilo, HI 96720, USA}
\affil{School of Astronomy and Space Science, Nanjing University, 163 Xianlin Avenue, Nanjing, Jiangsu 210023, People's Republic of China}
\affil{Key Laboratory of Modern Astronomy and Astrophysics (Nanjing University), Ministry of Education, Nanjing, Jiangsu 210023, People's Republic of China}
\affil{Center for Astrophysics $\vert$ Harvard \& Smithsonian, 60 Garden Street, Cambridge, MA 02138, USA}

\author[0000-0002-5093-5088]{Keping Qiu}
\affil{School of Astronomy and Space Science, Nanjing University, 163 Xianlin Avenue, Nanjing, Jiangsu 210023, People's Republic of China}
\affil{Key Laboratory of Modern Astronomy and Astrophysics (Nanjing University), Ministry of Education, Nanjing, Jiangsu 210023, People's Republic of China}

%\author{coauthors}
%\affil{Affli}
\author[0000-0003-2384-6589]{Qizhou Zhang}
\affil{Center for Astrophysics $\vert$ Harvard \& Smithsonian, 60 Garden Street, Cambridge, MA 02138, USA}

%\author{et al.}

\begin{abstract}
The Davis-Chandrasekhar-Fermi (DCF) method provides an indirect way to estimate the magnetic field strength from statistics of magnetic field orientations. We compile all the previous DCF estimations from polarized dust emission observations and re-calculate the magnetic field strength of the selected samples with the new DCF correction factors in \citet{2021ApJ...919...79L}. We find the magnetic field scales with the volume density as $B \propto n^{0.57}$. However, the estimated power-law index of the observed $B-n$ relation has large uncertainties and may not be comparable to the $B-n$ relation of theoretical models. A clear trend of decreasing magnetic viral parameter (i.e., increasing mass-to-flux ratio in units of critical value) with increasing column density is found in the sample, which suggests the magnetic field dominates the gravity at lower densities but cannot compete with the gravity at higher densities. This finding also indicates that the magnetic flux is dissipated at higher column densities due to ambipolar diffusion or magnetic reconnection, and the accumulation of mass at higher densities may be by mass flows along the magnetic field lines. Both sub-Alfv\'{e}nic and super-Alfv\'{e}nic states are found in the sample, with the average state being approximately trans-Alfv\'{e}nic. 
\end{abstract}
%\textbf{We suggest that it might be unnecessary to consider the cross-term magnetic field in molecular dense structures where self-gravity is important.}

\section{Introduction}
Two major issues, the relative importance between the magnetic field and gravity, and the relative importance between the magnetic field and turbulence, are still not yet solved in studies of magnetic fields in star formation. Observationally determining the magnetic field strength is crucial to address the two issues. 

Zeeman line splitting observations are the only way to directly measure the line-of-sight magnetic field strength. Previous Zeeman observations suggest that the magnetic support dominates gravity at lower densities, while gravity dominates magnetic support at higher densities \citep{2010ApJ...725..466C, 2012ARAandA..50...29C, 2019FrASS...6...66C}. These Zeeman studies also found that the magnetic field scales with density as $B \propto n^{2/3}$ at $n_{\mathrm{H_2}} > 150$ cm$^{-3}$ with Bayesian analysis, and proposed that the magnetic field is too weak to compete with gravity at high densities. However, the Zeeman observations only give the upper limit of the field strength, and the Bayesian analysis may be problematic \citep{2021Galax...9...41L}. Thus, independent estimations of the field strength with other methods are required to draw a more complete conclusion. 

The Davis-Chandrasekhar-Fermi (DCF) method \citep{1951PhRv...81..890D, 1953ApJ...118..113C} provides a powerful way to indirectly estimate the plane-of-sky magnetic field strength from observations of the polarized emission of dust grains that are aligned with the magnetic field. \citet{2019FrASS...6...15P} collected the magnetic field strength estimated from single-dish dust polarization observations with the DCF method and found that the $B-n$ relation from DCF estimations was comparable to that of the Zeeman observations. A DCF study of 17 molecular dense cores found a similar $B-n$ relation to those of the Zeeman studies \citep{2021ApJ...917...35M}. They also found that the cores are mildly magnetically super-critical and nearly trans-Alfv\'{e}nic on average. However, the conclusions of the two studies are limited by the sample size and the value range of physical parameters. It is essential to carry out a complete study of all the previous DCF estimations, which will greatly improve the statistics and extend our understandings of the magnetic field in star formation. 

Here we present a compilation of all the previous DCF estimations published in the literature. In Section \ref{sec:samana}, we demonstrate our sample selection and how we treat and re-calculate the physical parameters of the selected sample. In Section \ref{sec:resdis}, we show and discuss the results on the comparison between the magnetic field and gravity and between the magnetic field and turbulence. Section \ref{section:conclusion} presents the conclusions. 

\section{Sample and analysis} \label{sec:samana}
\subsection{Sample}
We collect all the previous DCF estimations using polarized dust emission observations. Similar to the approaches in \citet{2019FrASS...6...15P} and \citet{2021ApJ...917...35M}, we adopt the referenced properties of the collected sources with slight adjustments. If the $n_{\mathrm{H}}$ is given instead of the volume density $n_{\mathrm{H_2}}$, the $n_{\mathrm{H_2}}$ is estimated with $n_{\mathrm{H_2}} = 0.5 n_{\mathrm{H}}$. For filaments, the equivalent radius r is estimated with the relation $V = 4/3 \pi r^3$, where $V$ is the volume. We simply adopt the mass reported in the original references and neglect the slight difference in values of the mean molecular weight. If a range of values are given for a parameter, we only adopt the average value. If the same value of parameters is adopted for different sources, we only report the average properties of different sources \citep[e.g., the sources in][]{2016ApJ...825L..15C}. The physical parameters of the collected sources are reported in Appendix \ref{app:dcfref}. 

\subsection{Recalculation of magnetic field strength}
The DCF method is based on assumptions that the turbulence is isotropic, there is equipartition between turbulent kinetic energy and turbulent magnetic energy, and the turbulent-to-ordered magnetic field strength ratio is probed by the angular dispersion of magnetic field orientations. The DCF estimations are subject to uncertainties depending on the validity of the DCF assumptions and various effects in the estimation of the angular dispersion. Many attempts have been devoted to correct for one or several of the uncertainties by modifying the original DCF formula and/or applying a correction factor. Most previous DCF studies estimated the uniform magnetic field strength. Recently, we derived the up-to-date correction factors for the uniform and total magnetic field strength derived from the original DCF formula and different modified forms of the DCF equation at clump and core scales\footnote{Following the nomenclature in \citet{2009ApJ...696..268Z}, we refer to a molecular clump as an entity of $\sim$ 1 pc and a dense core as an entity of $\sim$ 0.1 pc} based on results of self-gravitating MHD and radiative transfer simulations \citep{2021ApJ...919...79L}. Using the physical parameters (density, turbulent velocity dispersion, angular dispersion, and turbulent-to-ordered field strength ratio) reported in the referenced DCF papers, here we recalculate the magnetic field strength of the same sources with the new correction factors in \citet{2021ApJ...919...79L}. Due to a lack of simulations with similar physical conditions to the large-scale regions observed by the Planck satellite, we do not apply any correction factors for the estimated field strength of the molecular clouds observed by the Planck. 
%adopted the original or modified DCF equations and estimated the uniform magnetic field strength with or without a correction factor

Other than the uncertainty of the DCF method itself, the uncertainties on the referenced physical properties (density, velocity dispersion) could also affect the reliability of the estimated magnetic field strength. The volume density is usually derived from dust thermal emission observations, which requires the information on the dust-to-gas ratio, temperature, dust opacity, and geometry of the studied source in question. The information on these parameters is not always available for the observed sources. Different works may have adopted different assumptions for these intermediate physical parameters, which means the volume densities in different works could be derived in inconsistent ways. On the other hand, the velocity dispersion is measured from spectral line observations. The particular line tracers used in different works have different chemical processes and excitation conditions, which can bring inconsistency in the derivation of the velocity dispersion. Moreover, the dust emission used to derive the density, the line emission used to derive the velocity dispersion, and the polarized dust emission used to derive the angular dispersion, should match each other in the sampled area, but many previous DCF studies did not check this consistency. Unfortunately, we are unable to rectify these consistency issues in this work because we do not have the dataset of the existing observations. We note that all these factors mentioned here could bring additional uncertainties in the analysis in this paper since we are conducting a statistical study of the existing observations. 

Recently, \citet{2021A&A...647A.186S} proposed a method that is similar to the DCF method and is based on a different assumption that there is an energy equipartition between the turbulent kinetic energy and the cross-term magnetic energy. We suggest that this assumption may not be valid in the commonly studied dense molecular structures where self-gravity is important (see Appendix \ref{app:cross} for a discussion). Thus, we do not apply this method on our sample to estimate the magnetic field strength.

\subsubsection{The original DCF method}

In the original DCF method \citep[hereafter DCF53, ][]{1951PhRv...81..890D, 1953ApJ...118..113C}, the plane-of-sky uniform magnetic field strength is estimated with\footnote{All the magnetic-related terms in this paper are in SI units or CGS units unless otherwise noted.}
\begin{equation}\label{eq:eqdcfrefsa}
B^{\mathrm{u,DCF53}}_{\mathrm{pos}} \sim \sqrt{\mu_0 \rho }\frac{\delta v_{\mathrm{los}}}{\delta\phi},
\end{equation}
where $\mu_0$ is the permeability of vacuum, $\rho$ is the gas density, $\delta v_{\mathrm{los}}$ is the line-of-sight turbulent velocity dispersion, and $\delta\phi$ is the measured angular dispersion. The gas density is estimated as $\rho = \mu_{\mathrm{H_2}} m_{\mathrm{H}} n_{\mathrm{H_2}}$, where $\mu_{\mathrm{H_2}}$ is the mean molecular weight per hydrogen molecule and $m_{\mathrm{H}}$ is the atomic mass of hydrogen. Many previous DCF studies adopted a correction factor of $Q_{c} \sim 0.5$ for the estimated uniform field strength based on the simulation results of \citet{2001ApJ...546..980O}. However, this correction factor is only applicable on low-density molecular clouds greater than pc scales. Recently, \citet{2021ApJ...919...79L} derived the correction factors of the original DCF equation for strong field models at clump and core scales that are with high density and significant self-gravity. Adopting the average correction factors for spherical and cylindrical structures in \citet{2021ApJ...919...79L}, the corrected plane-of-sky uniform magnetic field strength is estimated as 
\begin{equation}\label{eq:eqdcfsa}
B^{\mathrm{u,DCF53,est}}_{\mathrm{pos}} \sim 0.28 \sqrt{\mu_0 \rho }\frac{\delta v_{\mathrm{los}}}{\delta\phi},
\end{equation}
and the corrected plane-of-sky total magnetic field strength is estimated as 
\begin{equation}\label{eq:eqdcftotsa}
B^{\mathrm{tot,DCF53,est}}_{\mathrm{pos}} \sim 0.62 \sqrt{\mu_0 \rho }\frac{\delta v_{\mathrm{los}}}{\delta\phi},
\end{equation}
We adopt $\mu_{\mathrm{H_2}} = 2.86$ \citep{2013MNRAS.432.1424K} in the derivation of the gas density. We only calculate the $B^{\mathrm{u,dcf53,est}}_{\mathrm{pos}}$ when $\delta\phi < 25\degr $ \citep{2021ApJ...919...79L}. 

\subsubsection{The modified DCF methods that considered the ordered field contribution}
Several works tried to remove the contribution from the ordered field structure to the measured angular dispersion. The attempts include fitting the field structure with a hourglass shape for strong field cases \citep[hereafter the HG method, e.g., ][]{2006Sci...313..812G},  the spatial filtering method \citep[hereafter the Pil15 method, ][]{2015ApJ...799...74P}, and the unsharp masking method \citep[hereafter the Pat17 method, ][]{2017ApJ...846..122P}. These works usually also adopt a correction factor of $Q_{c} \sim 0.5$ for their estimated field strength. Similarly, we recalculate the field strength for the ordered-structure-removal methods (hereafter the OSR methods). \citet{2021ApJ...919...79L} did not explicitly report the correction factors for the OSR methods, but shows that the contribution from ordered field structure can overestimate the angular dispersion by a factor of $\sim$2.5 on average. Thus, we divide the correction factors of the DCF53 method by a factor of 2.5 to obtain the correction factors for the OSR methods. Thus, the corrected plane-of-sky uniform magnetic field strength is estimated as 
\begin{equation}\label{eq:eqosr}
B^{\mathrm{u,OSR,est}}_{\mathrm{pos}} \sim 0.11 \sqrt{\mu_0 \rho }\frac{\delta v_{\mathrm{los}}}{\delta\phi},
\end{equation}
and the corrected plane-of-sky total magnetic field strength is estimated as 
\begin{equation}\label{eq:eqosrtot}
B^{\mathrm{tot,OSR,est}}_{\mathrm{pos}} \sim 0.25 \sqrt{\mu_0 \rho }\frac{\delta v_{\mathrm{los}}}{\delta\phi}.
\end{equation}

\subsubsection{The angular dispersion function method}

The angular dispersion function method \citep[hereafter the ADF method. ][]{2008ApJ...679..537F, 2009ApJ...696..567H, 2009ApJ...706.1504H, 2016ApJ...820...38H} analytically accounts for various of effects that may affect the measured angular dispersion. Based on different effects considered, the ADF methods can be divided into the structure function method \citep[hereafter the Hil09 method, ][]{2009ApJ...696..567H}, the auto-correlation function method for single-dish observations \citep[hereafter the Hou09 method, ][]{2009ApJ...706.1504H}, and the auto-correlation function method for interferometer observations \citep[hereafter the Hou16 method, ][]{2016ApJ...820...38H}. Similarly, we adopt the correction factors in \citet{2021ApJ...919...79L} for strong field models. For the Hil09 method, the corrected plane-of-sky uniform magnetic field strength is estimated as 
\begin{equation}\label{eq:eqhil09u}
B^{\mathrm{u,Hil09,est}}_{\mathrm{pos}} \sim 0.1 \sqrt{\mu_0 \rho }\delta v_{\mathrm{los}}((\frac{\langle B_{\mathrm{t}}^2 \rangle}{ \langle B_0^2\rangle})^{-\frac{1}{2}})^{\mathrm{Hil09}}
\end{equation}
and the corrected plane-of-sky total magnetic field strength is estimated as 
\begin{equation}\label{eq:eqhil09tot}
B^{\mathrm{tot,Hil09,est}}_{\mathrm{pos}} \sim 0.21 \sqrt{\mu_0 \rho }\delta v_{\mathrm{los}}((\frac{\langle B_{\mathrm{t}}^2 \rangle}{ \langle B^2\rangle})^{-\frac{1}{2}})^{\mathrm{Hil09}},
\end{equation}
where $((\langle B_{\mathrm{t}}^2 \rangle/  \langle B_0^2\rangle)^{0.5})^{\mathrm{Hil09}}$ is the turbulent-to-ordered field strength ratio and $((\langle B_{\mathrm{t}}^2 \rangle/  \langle B^2\rangle)^{0.5})^{\mathrm{Hil09}}$ is the turbulent-to-total field strength ratio derived from this method. We do not calculate $B^{\mathrm{u,Hil09,est}}_{\mathrm{pos}}$ if $((\langle B_{\mathrm{t}}^2 \rangle/  \langle B_0^2\rangle)^{0.5})^{\mathrm{Hil09}}>0.1$ \citep{2021ApJ...919...79L}. For the Hou09 and Hou16 methods, the corrected plane-of-sky uniform magnetic field strength is estimated as 
\begin{equation}\label{eq:eqhouu}
B^{\mathrm{u,Hou,est}}_{\mathrm{pos}} \sim 0.19 \sqrt{\mu_0 \rho }\delta v_{\mathrm{los}}((\frac{\langle B_{\mathrm{t}}^2 \rangle}{ \langle B_0^2\rangle})^{-\frac{1}{2}})^{\mathrm{Hou}}
\end{equation}
and the corrected plane-of-sky total magnetic field strength is estimated as 
\begin{equation}\label{eq:eqhoutot}
B^{\mathrm{tot,Hou,est}}_{\mathrm{pos}} \sim 0.39 \sqrt{\mu_0 \rho }\delta v_{\mathrm{los}}((\frac{\langle B_{\mathrm{t}}^2 \rangle}{ \langle B^2\rangle})^{-\frac{1}{2}})^{\mathrm{Hou}},
\end{equation}
where $((\langle B_{\mathrm{t}}^2 \rangle/  \langle B_0^2\rangle)^{0.5})^{\mathrm{Hou}}$ is the turbulent-to-ordered field strength ratio and $((\langle B_{\mathrm{t}}^2 \rangle/  \langle B^2\rangle)^{0.5})^{\mathrm{Hou}}$ is the turbulent-to-total field strength ratio derived from the two methods. Due to the limitation that the angular dispersion cannot exceed the value expected for a random field, the maximum derivable turbulent-to-ordered field strength ratio from the ADF methods is 0.76 \citep{2021ApJ...919...79L}. Thus, if the derived $((\langle B_{\mathrm{t}}^2 \rangle/  \langle B_0^2\rangle)^{0.5})^{\mathrm{Hou}}$ is greater than 0.76, the assumptions underlying the ADF methods on accounting for the line-of-sight signal integration may not be valid, which could lead to overestimation of the turbulent-to-ordered field strength ratio and underestimation of the field strength. In this situation, we adopt the turbulent-to-ordered field strength ratio $((\langle B_{\mathrm{t}}^2 \rangle/  \langle B_0^2\rangle)^{0.5})^{\mathrm{Hou}}_{\mathrm{nosi}}$ and the turbulent-to-total field strength ratio $((\langle B_{\mathrm{t}}^2 \rangle/  \langle B^2\rangle)^{0.5})^{\mathrm{Hou}}_{\mathrm{nosi}}$ without accounting for the line-of-sight signal integration and apply the same correction factors as those for the Hil09 method. Then the corrected plane-of-sky uniform magnetic field strength is estimated as 
\begin{equation}\label{eq:eqhouumod}
B^{\mathrm{u,Hou,est}}_{\mathrm{pos}} \sim 0.1 \sqrt{\mu_0 \rho }\delta v_{\mathrm{los}}((\frac{\langle B_{\mathrm{t}}^2 \rangle}{ \langle B_0^2\rangle})^{-\frac{1}{2}})^{\mathrm{Hou}}_{\mathrm{nosi}}
\end{equation}
and the corrected plane-of-sky total magnetic field strength is estimated as 
\begin{equation}\label{eq:eqhoutotmod}
B^{\mathrm{tot,Hou,est}}_{\mathrm{pos}} \sim 0.21 \sqrt{\mu_0 \rho }\delta v_{\mathrm{los}}((\frac{\langle B_{\mathrm{t}}^2 \rangle}{ \langle B^2\rangle})^{-\frac{1}{2}})^{\mathrm{Hou}}_{\mathrm{nosi}},
\end{equation}
where $((\langle B_{\mathrm{t}}^2 \rangle/  \langle B_0^2\rangle)^{0.5})^{\mathrm{Hou}}_{\mathrm{nosi}} = ((\langle B_{\mathrm{t}}^2 \rangle/  \langle B_0^2\rangle)^{0.5})^{\mathrm{Hou}} / \sqrt{N_{\mathrm{adf}}}$. $N_{\mathrm{adf}}$ is the number of turbulent fluid elements along the line of sight. Similarly, we do not calculate $B^{\mathrm{u,Hou,est}}_{\mathrm{pos}}$ if $((\langle B_{\mathrm{t}}^2 \rangle/  \langle B_0^2\rangle)^{0.5})^{\mathrm{Hou}}_{\mathrm{nosi}} >0.1$ \citep{2021ApJ...919...79L}.

\subsubsection{The \citet{2001ApJ...561..800H} method}

\citet{2001ApJ...561..800H} proposed that the plane-of-sky or 3D uniform magnetic field strength is estimated with
\begin{equation}\label{eq:eqhei01u}
B^{\mathrm{u,Hei01}}_{\mathrm{pos/3D}} \sim \sqrt{\mu_0 \rho }\frac{\delta v_{\mathrm{los}}}{\delta (\tan \phi)}
\end{equation}
if the turbulent magnetic energy is isotropic and the mean magnetic field orientation is along the plane of sky. They also proposed that the 3D total magnetic field strength is estimated as
\begin{equation}\label{eq:eqhei01tot}
B^{\mathrm{tot,Hei01}}_{\mathrm{3D}} \sim \sqrt{\mu_0 \rho }\frac{\delta v_{\mathrm{los}}}{\delta (\tan \phi)}(1+3\delta (\tan \phi)^2)^{1/2}
\end{equation} 
with the same assumptions. We neglect the magnetic field strength estimations with this method because of the large uncertainty associated with the scatters of $\delta (\tan \phi)$ \citep{2021ApJ...919...79L}.

\subsubsection{The \citet{2008ApJ...679..537F} method}

\citet{2008ApJ...679..537F} proposed that the plane-of-sky total magnetic field strength is estimated as (hereafter the Fal08 equation)
\begin{equation}\label{eq:eqFal08tot}
B^{\mathrm{tot,Fal08}}_{\mathrm{pos}} \sim \sqrt{\mu_0 \rho }\frac{\delta v_{\mathrm{los}}}{\tan (\delta \phi)}.
\end{equation} 
However, this formula is only valid in small angle approximation where $\tan (\delta \phi) \sim \delta \phi$ and $B^{\mathrm{tot}} \sim B^{\mathrm{u}}$ \citep{2021ApJ...919...79L}, then the formula would be identical to the original DCF equation. Thus we just recalculate the magnetic field strength with Equations \ref{eq:eqdcfsa} and \ref{eq:eqdcftotsa} for the Fal08 method. 

\subsubsection{Comparison between referenced and our estimated field strength}

\begin{figure*}[!htbp]
 \gridline{\fig{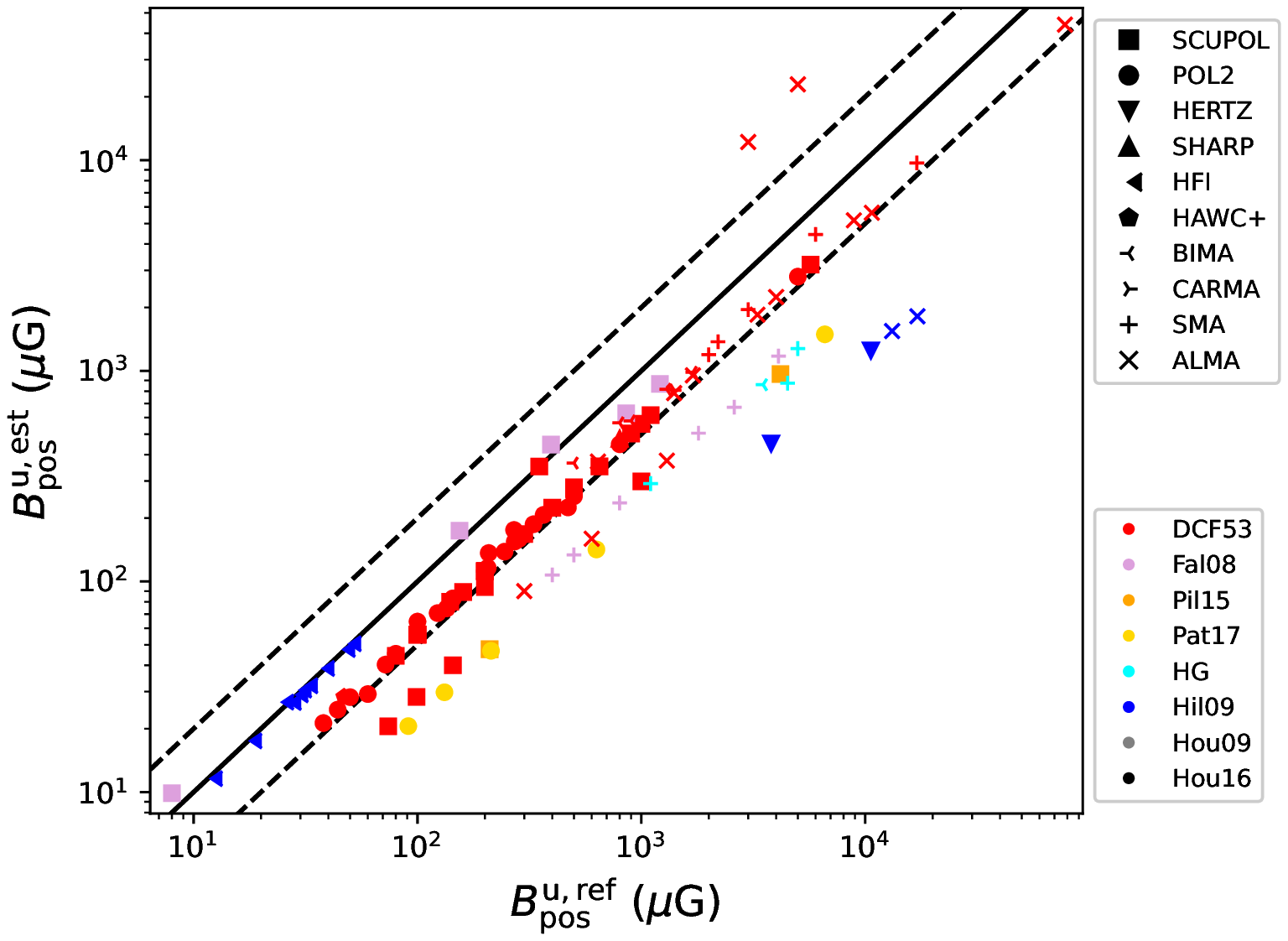}{0.45\textwidth}{(a)}
 \fig{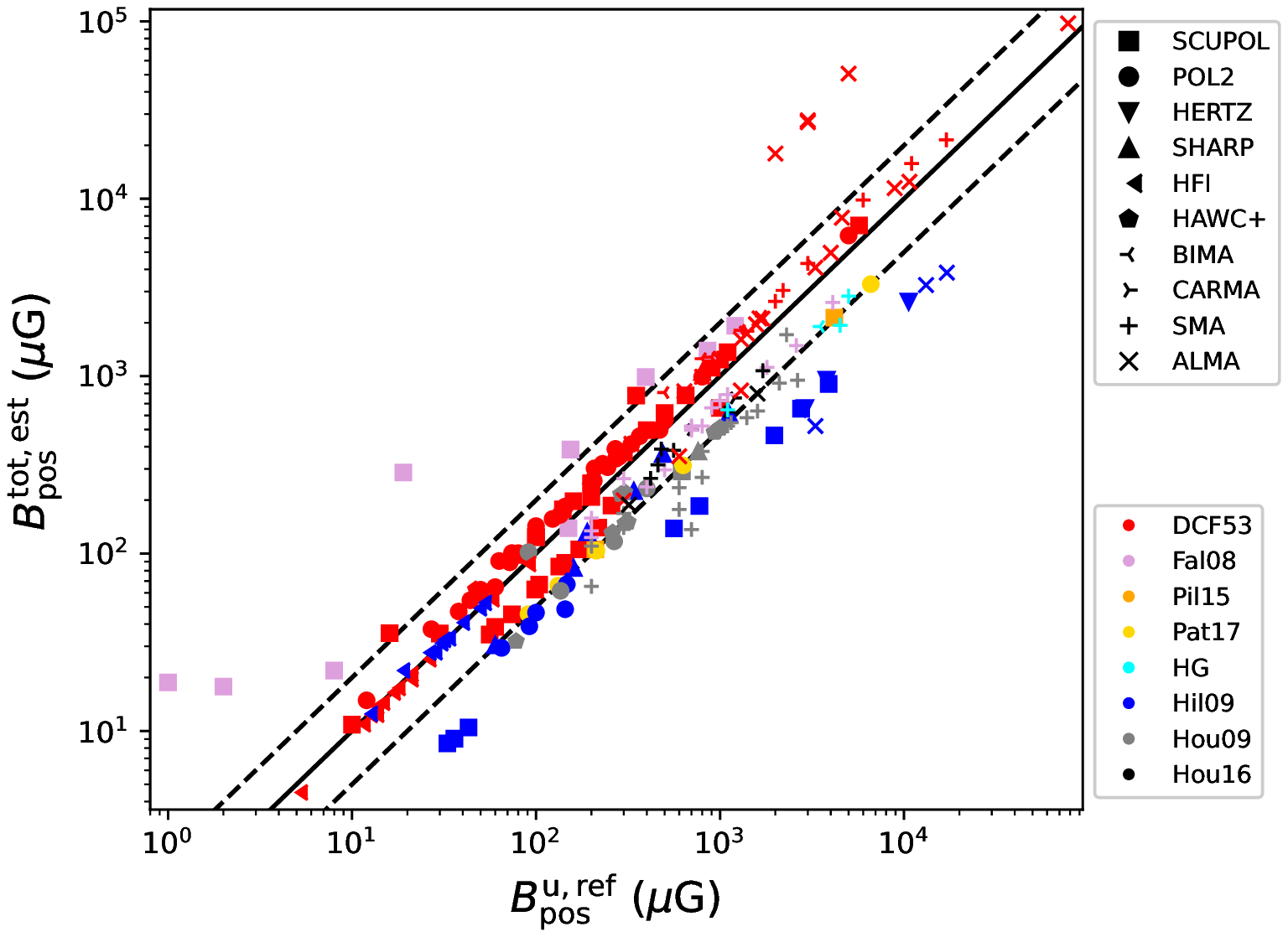}{0.45\textwidth}{(b)}
 }
\caption{(a). Compare the re-estimated and referenced uniform magnetic field strength. (b). Compare our estimated total magnetic field strength with the referenced uniform magnetic field strength. Different symbols represent different instruments. Different colors represent different methods. Solid lines correspond
to 1:1 relation. Dashed lines correspond to 1:2 and 2:1 relations.  \label{fig:B_B}}
\end{figure*}

Figure \ref{fig:B_B}(a) shows a comparison between our re-estimated plane-of-sky uniform magnetic field strength $B^{\mathrm{u,est}}_{\mathrm{pos}}$ and the referenced plane-of-sky uniform magnetic field strength $B^{\mathrm{u,ref}}_{\mathrm{pos}}$. Generally $B^{\mathrm{u,est}}_{\mathrm{pos}}$ is smaller than $B^{\mathrm{u,ref}}_{\mathrm{pos}}$, which is because of the smaller correction factors adopted in our recalculations. On the other hand, Figure \ref{fig:B_B}(b) shows that our estimated plane-of-sky total magnetic field strength $B^{\mathrm{tot,est}}_{\mathrm{pos}}$ is generally comparable to $B^{\mathrm{u,ref}}_{\mathrm{pos}}$.

\section{Results and Discussions} \label{sec:resdis}
\subsection{$n - r$ relation}\label{sec:nr}

\begin{figure}[!htbp]
 \gridline{\fig{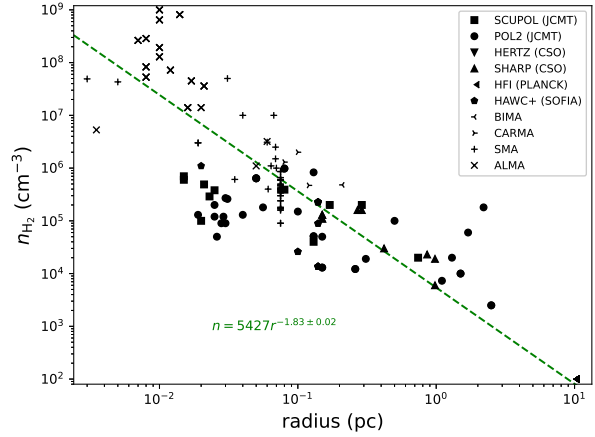}{0.45\textwidth}{}}
\caption{The relation between density $n$ and radius $r$ for the selected sample. Different symbols represent different instruments. The dashed line is the result of a least-squares fit. \label{fig:n_r}}
\end{figure}

In Figure \ref{fig:n_r}, we plot the relation between the volume density $n$ and the radius $r$ for sources with estimations of both $n$ and $r$. We fit the $n - r$ relation with a simple \textbf{power-law $n \propto r^{-i}$}. The derivation of the radius is usually straightforward and associated with small uncertainties for compact sources, so we fix the radius in the $n - r$ fitting. Assuming an uncertainty of a factor of 2 for the volume density, the power-law index $i$ is estimated \textbf{to be $1.83 \pm 0.02$}. The density profiles for a self-gravitating structure in free-fall and in hydrostatic equilibrium are $n \propto r^{-1.5}$ and $n \propto r^{-2}$, respectively  \citep{1987ARAandA..25...23S}. Our estimated power-law index is between the values predicted by the two models. Note that our collected $n - r$ relation is based on a collection of different sources, while the $n - r$ relation for free-fall or hydrostatic equilibrium models is based on the density profile of a single structure. 

\subsection{Comparing magnetic field with gravity}
\subsubsection{$B - n$ relation}

The power-law index of the $B - n$ relation ($B \propto n^{j}$) can be used as a test for the dynamical importance of the magnetic field during the compression of dense gas structures \citep{2012ARAandA..50...29C}. For a dense core where the magnetic field is too weak to prevent isotropic gas collapse, a power-law index of $2/3$ is predicted \citep{1966MNRAS.133..265M}. For a strong field case where ambipolar diffusion is expected to be the dominant mechanism for the gas collapse, the magnetic field would increase faster than the density during gas compression, resulting in a power-law index of $<0.5$ \citep{1999ASIC..540..305M}. 

\begin{figure*}[!htbp]
 \gridline{\fig{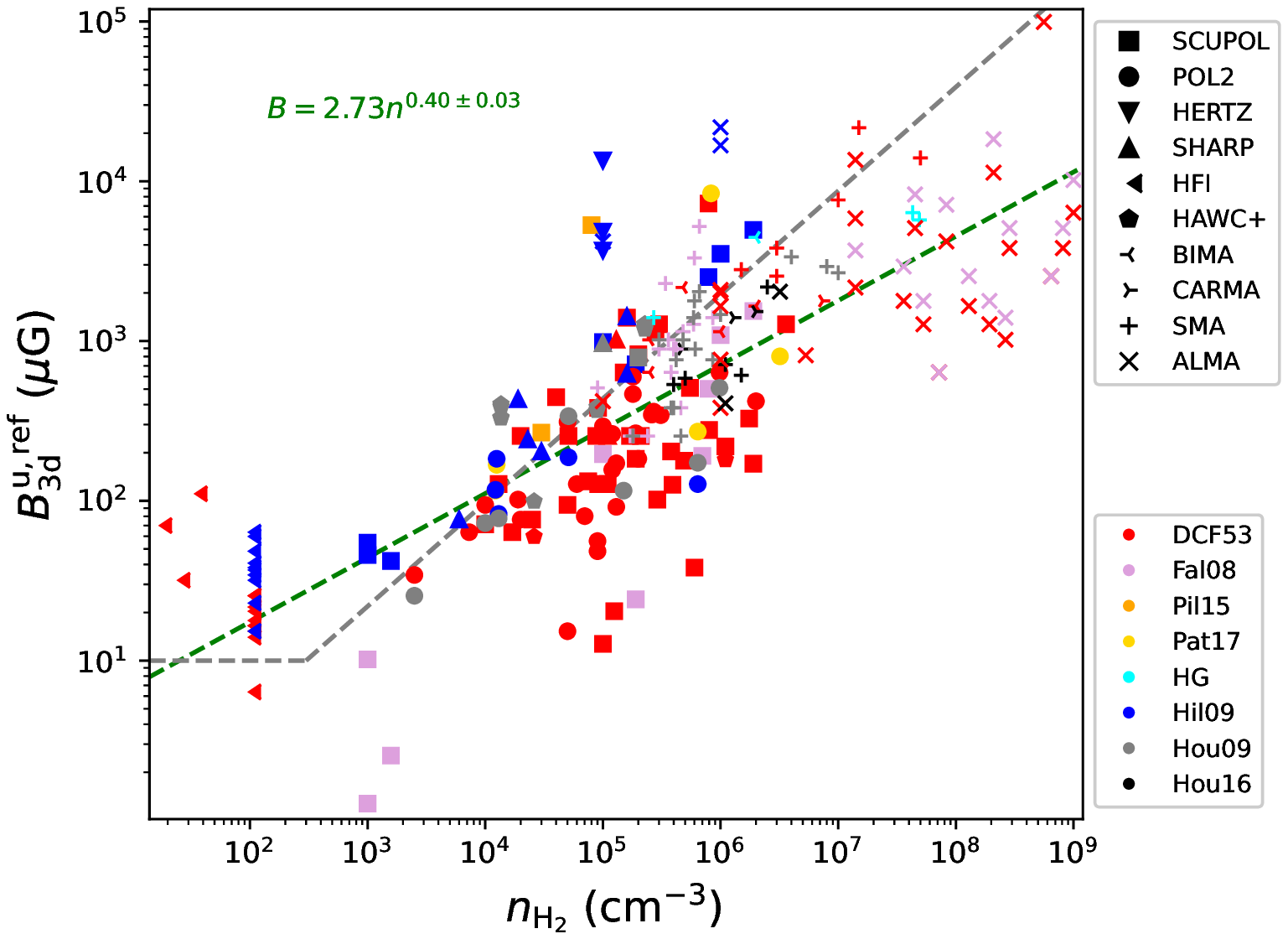}{0.45\textwidth}{(a)}
 \fig{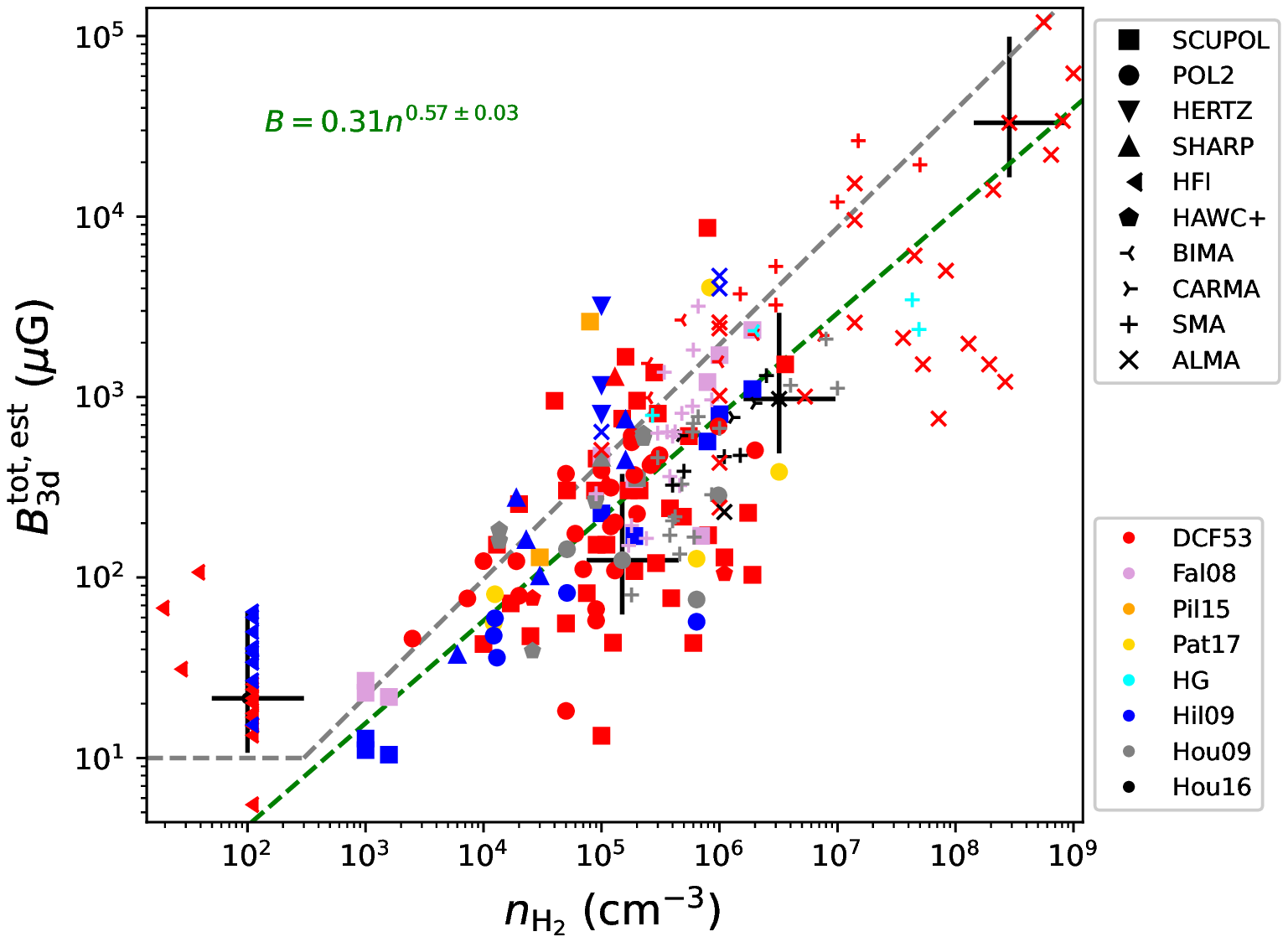}{0.45\textwidth}{(b)}
 }
\caption{(a). Relation between the referenced 3D uniform magnetic field strength and the volume density. (b). Relation between our estimated 3D total magnetic field strength and the volume density. Example errorbars for uncertainties of a factor of 2 are shown for several sources at different densities. Different symbols represent different instruments. Different colors represent different methods. The green dashed lines are the results of least-square fits. The grey dashed lines mark the relation found in Zeeman observations \citep{2010ApJ...725..466C} for comparison.  \label{fig:B_n}}
\end{figure*}

The referenced plane-of-sky uniform magnetic field strength is converted to the 3D uniform field strength with the statistical relation $B^{\mathrm{u,ref}}_{\mathrm{3D}} = \frac{4}{\pi}B^{\mathrm{u,ref}}_{\mathrm{pos}}$ \citep{2004ApJ...600..279C}. Figure \ref{fig:B_n}(a) shows the relation between $B^{\mathrm{u,ref}}_{\mathrm{3D}}$ and $n_{\mathrm{H_2}}$ in logarithmic scales. Assuming uncertainties of a factor of 2 for both the field strength and the volume density, we derive a power-law index of $0.40 \pm 0.03$ with a simple power-law fit $B \propto n^{j}$. Due to clear deviations from the general power-law trend, data points from Planck observations are not considered in the fitting. Because the field strength estimations in the referenced papers might not adopt the appropriate correction factors and the conversion from plane-of-sky uniform field strength to 3D uniform field strength could have large uncertainties \citep{2021ApJ...919...79L}, we refrain from further interpreting the derived power-law index from the referenced $B - n$ relation.
%the derived power-law index from the referenced $B - n$ relation may not be appropriate to be compared with the values predicted by theoretical models. 

We re-investigate the $B - n$ relation from our new estimations. We convert our estimated plane-of-sky total magnetic field strength to the 3D total field strength with the statistical relation $B^{\mathrm{tot,est}}_{\mathrm{3D}} = \sqrt{\frac{3}{2}}B^{\mathrm{tot,est}}_{\mathrm{pos}}$ \citep{2021ApJ...919...79L} and show the relation between $B^{\mathrm{tot,est}}_{\mathrm{3D}}$ and $n_{\mathrm{H_2}}$ in Figure \ref{fig:B_n}(b). The conversion from plane-of-sky field strength to 3D field strength for the total field has less uncertainties than the uniform field and is thus more reliable \citep{2021ApJ...919...79L}. Similarly, we fit the relation with a simple power-law $B \propto n^{j}$ and derive a power-law index of $0.57 \pm 0.03$ by assuming uncertainties of a factor of 2 for both parameters. Data points from Planck observations are not considered in the fitting as well. 

Our derived power-law index of 0.57 is between the value expected for a strong field model ($j \lesssim 0.5$) and for a weak field model ($j\sim2/3$). The estimated index is shallower than the value of 0.65 found by previous Zeeman observations \citep{2010ApJ...725..466C}, which might be due to the different physical conditions traced by Zeeman observations and dust polarization observations \citep{2013ApJ...777..112P}. The $B \propto n^{0.5}$ dependence of the DCF method may bias the power-law index toward $j=0.5$ (see discussions below in the same section), which might also be responsible for our lower $j$ value compared to the Zeeman results. Our estimated index is also different from the value of 0.66 found by a collection of DCF estimations in 17 dense cores \citep{2021ApJ...917...35M}, which might be due to their much smaller sample size.
%, which might imply that the magnetic field plays a moderate role in the contraction and collapse of star-forming dense structures

As addressed in \citet{2021Galax...9...41L}, there could be some problems when comparing the observed $B - n$ relation with the $B - n$ relation of theoretical models. Firstly, the observed $B - n$ relation is based on a collection of different sources, while the $B - n$ relation of theoretical models is based on the temporal evolution of a single structure. Also note that the temporal $B - n$ index may not be a constant of time \citep{2021Galax...9...41L}. Therefore, it is still under debate whether the observed $B - n$ relation is comparable to the model $B - n$ relation.

The investigation of the $B - n$ relation using DCF estimations could be limited by the $B \propto \sqrt{n}$ dependence of the DCF method. If we divide both $B$ and $n$ by $\sqrt{n}$, the $B - n$ relation is reduced to the relation between $\delta v_{\mathrm{los}} (B_{\mathrm{t}}/B)^{-1}$ and $\sqrt{n}$. i.e., $B \propto n^{0.5+j'} $, where $\delta v_{\mathrm{los}} (B_{\mathrm{t}}/B)^{-1} \propto n^{\frac{j'}{2}}$. However, the volume density spans over seven orders of magnitude in our sample, while the velocity dispersion and the turbulent-to-total field strength ratio (whether derived from ADF method or in the form of angular dispersions) span only one order of magnitude (see Appendix \ref{app:dcfref}). This could limit the sensitivity of using the $\delta v_{\mathrm{los}} (B_{\mathrm{t}}/B)^{-1} - \sqrt{n}$ relation to derive $j'$. Thus, the derived $B - n$ relation might be mostly determined by the $B \propto \sqrt{n}$ dependence of the DCF formula. 
% This is because the two variables $B$ and $n$ are not independent if the $B$ strength is derived from the DCF method. The DCF formula of $B$ explicitly contains the expression $\sqrt{n}$.
%When we investigate the Statistical relation between the two variables $B$ and $n$, we would want the two variables to be independent.

\begin{figure}[!htbp]
 \gridline{\fig{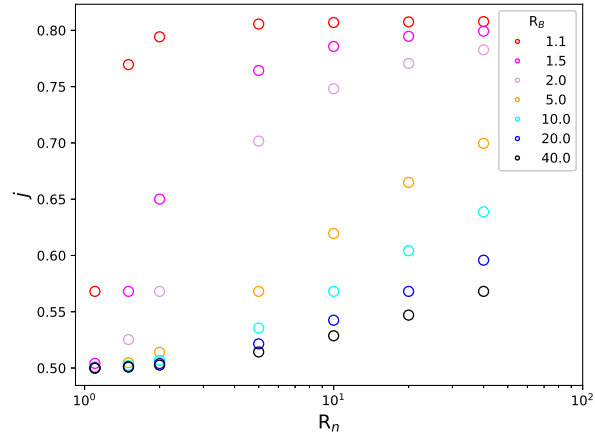}{0.45\textwidth}{}}
\caption{The fitted power-law index $j$ of the $B - n$ relation under different uncertainties of volume density and magnetic field strength R$_n$ and R$_B$. \label{fig:i_rn}}
\end{figure}

The large scatters of the $B - n$ data points and the uncertainties associated with $B$ and $n$ could also be problematic. As shown in Figure \ref{fig:B_n}(b), although the $B - n$ relation shows a clear power-law, the magnetic field strength can span up to 2 orders of magnitude at a give volume density. Thus, the fitted power-law index could be very unreliable, especially when we are only assuming an uncertainty of a factor of 2 (R=2) for both $B$ and $n$. Previous studies have shown that the uncertainties of the density and magnetic field strength estimates could be larger than a factor of 2 \citep{2015MNRAS.451.4384T, 2020ApJ...890..153J, 2021ApJ...919...79L}. We vary the uncertainties of the volume density (R$_n$) and the magnetic field (R$_B$) from 1.1 to 40 and re-fit the data to derive $j$ for each pairs of R$_n$ and R$_B$. The results are shown in Figure \ref{fig:i_rn}. We can see from Figure \ref{fig:i_rn} that $j=0.57$ is only obtained when R$_n$ $\sim$ R$_B$. When R$_n$ and R$_B$ are different, even only for R$<2$, the fitted $j$ can vary from 0.5 to 0.8 for our data, which makes it impossible to compare the derived $j$ with a strong field model ($j \lesssim 0.5$) or a weak field model ($j\sim2/3$). Thus, we conclude that we cannot give a reliable estimate of $j$ for our sample when the uncertainties of $n$ and $B$ are unknown.
%If different uncertainties for $B$ and $n$ are assumed, the derived power-law index could be very different. e.g., If we assume the uncertainties for the density and the magnetic field strength are a factor of 2 and 5, respectively, the fitted power-law index would be 0.5, which agrees with a strong field scenario. If the uncertainties for the density and the magnetic field strength are a factor of 5 and 2, respectively, the fitted power-law index would be 0.7, which approximately agrees with a weak field scenario.

\subsubsection{$B - N$ relation}

The relation between the magnetic field $B$ and the column density $N_{\mathrm{H_2}}$ can be used to investigate the relative importance between magnetic field and gravity. Based on the virial theorem, the magnetic virial mass is given by \citep{2020ApJ...895...142L}
\begin{equation}
M_{\mathrm{B}} = \frac{\pi R^2 B}{\sqrt{\frac{3}{2k}\mu_0\pi G}},
\end{equation}
where $R$ is the equivalent radius of a dense structure, $G$ is the gravitational constant, and $k = (5-2i)/(3-i)$ is a correction factor for the \textbf{density profile $n \propto r^{-i}$}. The relative importance between magnetic field and gravity can be stated with the magnetic virial parameter
\begin{equation}
\alpha_{\mathrm{B}} = \frac{M_{\mathrm{B}}}{M} = \frac{1}{\mu_{\mathrm{H_2}} m_{\mathrm{H}}} \sqrt{\frac{2k}{3\mu_0\pi G}} \frac{B}{N_{\mathrm{H_2}}},
\end{equation}
where $\mu_{\mathrm{H_2}} = 2.86$ is the mean molecular weight per hydrogen molecule \citep{2013MNRAS.432.1424K}, $m_{\mathrm{H}}$ is the atomic mass of hydrogen, and $M = \pi R^2 N_{\mathrm{H_2}}$ is the mass. Alternatively, the magnetic field and the gravity can also be compared using the magnetic critical parameter (i.e., mass-to-flux-ratio in units of its critical value)
\begin{equation}
\lambda_{\mathrm{B}} = \frac{1}{\alpha_{\mathrm{B}}} = \frac{M}{M_{\mathrm{B}}} = \mu_{\mathrm{H_2}} m_{\mathrm{H}} \sqrt{\frac{3\mu_0\pi G}{2k}} \frac{N_{\mathrm{H_2}}}{B}.
\end{equation}
When $i=1$, the form of $\lambda_{\mathrm{B}}$ would be identical to that of Equation 1 in \citet{2004ApJ...600..279C}. When $\alpha_{\mathrm{B}} < 1$ (i.e., $\lambda_{\mathrm{B}} > 1$, magnetically sub-virial, or magnetically super-critical), the gravity is stronger than the magnetic field, and vice versa. Both $\alpha_{\mathrm{B}}$ and $\lambda_{\mathrm{B}}$ can be reduced to the ratio between the magnetic field strength and the column density. It should also be noted that the $n - r$, $B - n$, and $B - N$ relations are not independent if a spherical structure is adopted ($N = 4nr/3$).

\begin{figure*}[!htbp]
 \gridline{\fig{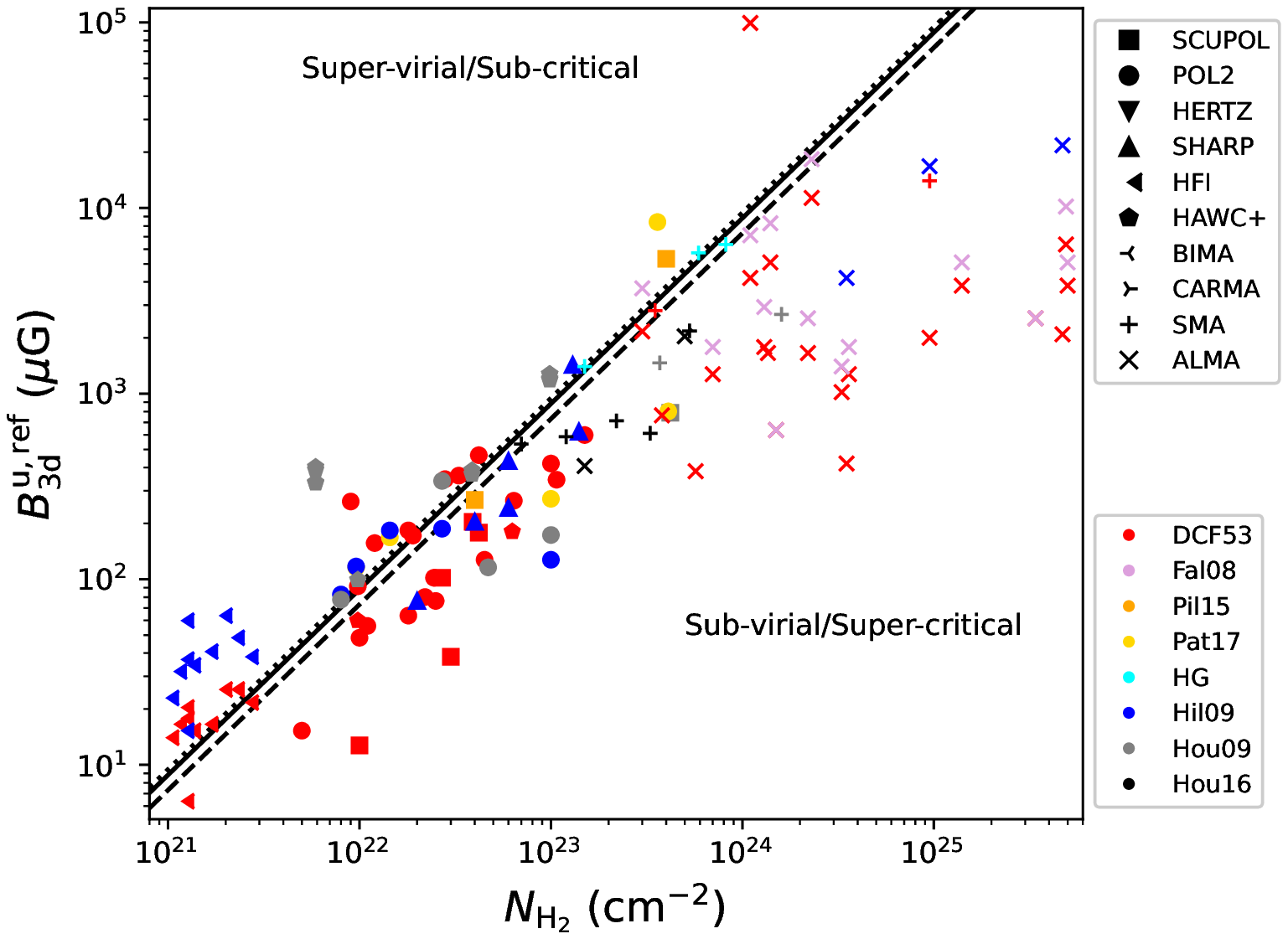}{0.45\textwidth}{}
 \fig{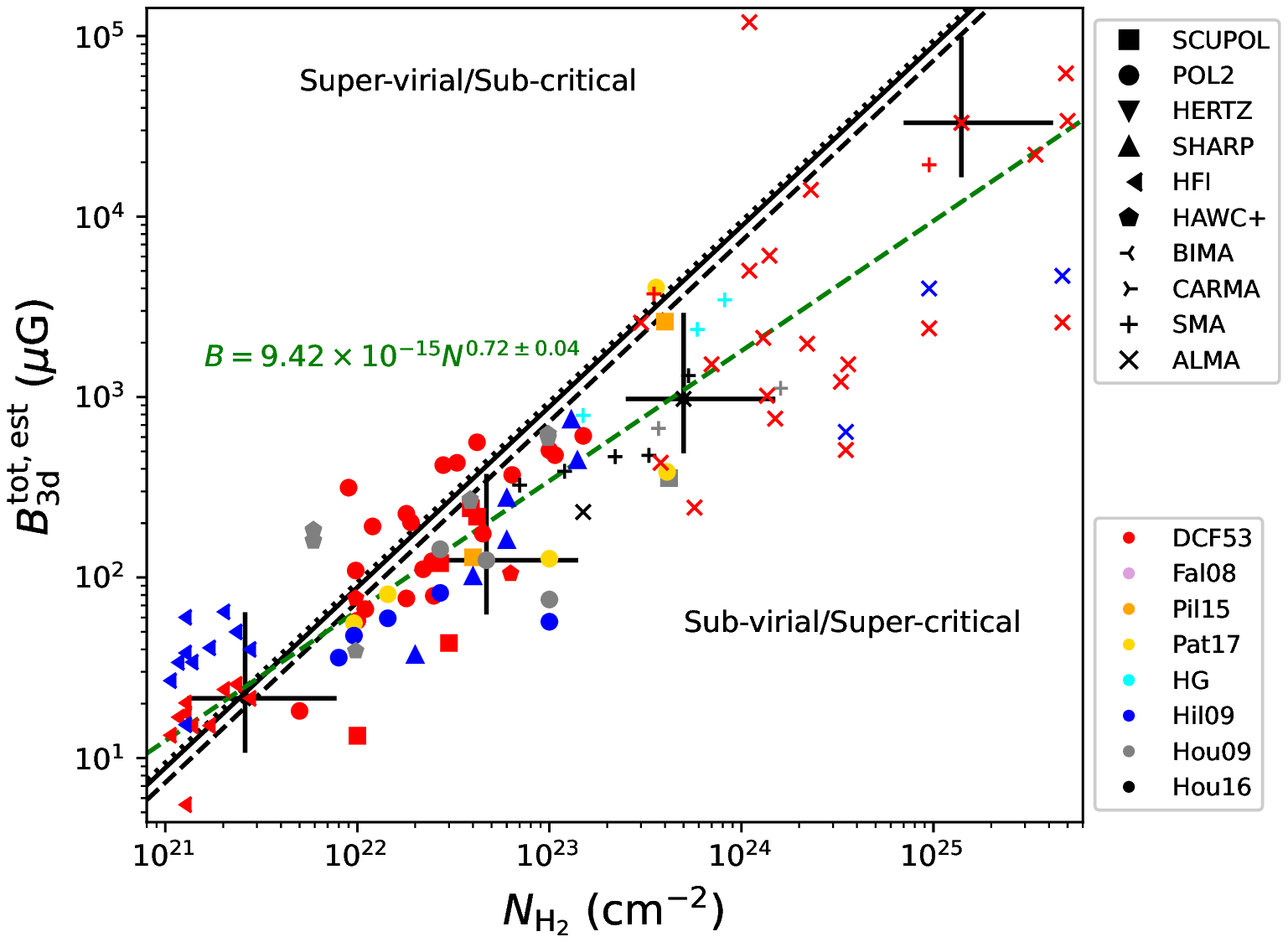}{0.45\textwidth}{}
 }
\caption{(a). Relation between the referenced 3D uniform magnetic field strength and the column density. (b). Relation between our estimated 3D total magnetic field strength and the column density. Example errorbars for uncertainties of a factor of 2 are shown for several sources at different densities. Different symbols represent different instruments. Different colors  represent different methods. The green dashed line is the result of a least-square fit. The black solid line corresponds to $\alpha_{\mathrm{B}} (i=1.8) = 1$, where the power-law index of $i=1.8$ is derived from the $n-r$ relation of our sample (see Section \ref{sec:nr}). The black dashed line corresponds to $\alpha_{\mathrm{B}} (i=0) = 1$ for structures with uniform density. The black dotted line corresponds to $\alpha_{\mathrm{B}} (i=2) = 1$ for structures in hydrostatic equilibrium.  \label{fig:B_N}}
\end{figure*}

Figure \ref{fig:B_N}(a) shows the relation between the referenced 3D uniform magnetic field strength $B^{\mathrm{u,ref}}_{\mathrm{3D}}$ and the column density $N_{\mathrm{H_2}}$. The trans-critical lines of $\alpha_{\mathrm{B}} = 1$ for $i=$0, 1.8, and 2 are shown in the figure. The structures at lower column density seem to be more super-virial/sub-critical, while the structures at higher column density tend to be more sub-virial/super-critical. However, because the field strength estimations might not adopt the appropriate correction factors and the contribution from turbulent field to the magnetic support were ignored in the referenced papers, this conclusion still needs further confirmation.

Thus, we re-investigate the $B - N$ relation from our new estimations. Figure \ref{fig:B_N}(b) shows the relation between our estimated 3D total magnetic field strength $B^{\mathrm{tot,est}}_{\mathrm{3D}}$ and the column density $N_{\mathrm{H_2}}$. At $N_{\mathrm{H_2}} < 3 \times 10^{21}$ cm$^{-2}$, the molecular clouds observed by Planck are nearly all sub-critical. At $3 \times 10^{21}$ cm$^{-2} < N_{\mathrm{H_2}} < 5 \times 10^{22}$ cm$^{-2}$, both sub- and super-critical molecular clumps/cores are found, with the average clumps/cores being slightly super-critical. For $N_{\mathrm{H_2}} > 5 \times 10^{22}$ cm$^{-2}$, nearly all the data points are super-critical. Assuming uncertainties of a factor of 2 for both $B$ and $N$, we derive a power-law index of $0.72 \pm 0.04$ with a simple power-law fit for the $B - N$ relation. Data points from Planck observations are not considered in the fitting. With the same reasons as for the $B - n$ index, the fitted $B - N$ index also has large uncertainties, so we refrain from further interpreting the $B - N$ index. But we notice that the derived power-law index is clearly smaller than 1, confirming that the gravity becomes more important as column density increases. The fitted $B - N$ line and the trans-critical line for $i=$1.8 intersect at $N_{\mathrm{H_2}} \sim 3.4 \times 10^{21}$ cm$^{-2}$ and $B^{\mathrm{tot,ref}}_{\mathrm{3D}} \sim 30 \mu$G, where the transition from sub-critical to super-critical occurs. 

\begin{figure}[!htbp]
 \gridline{\fig{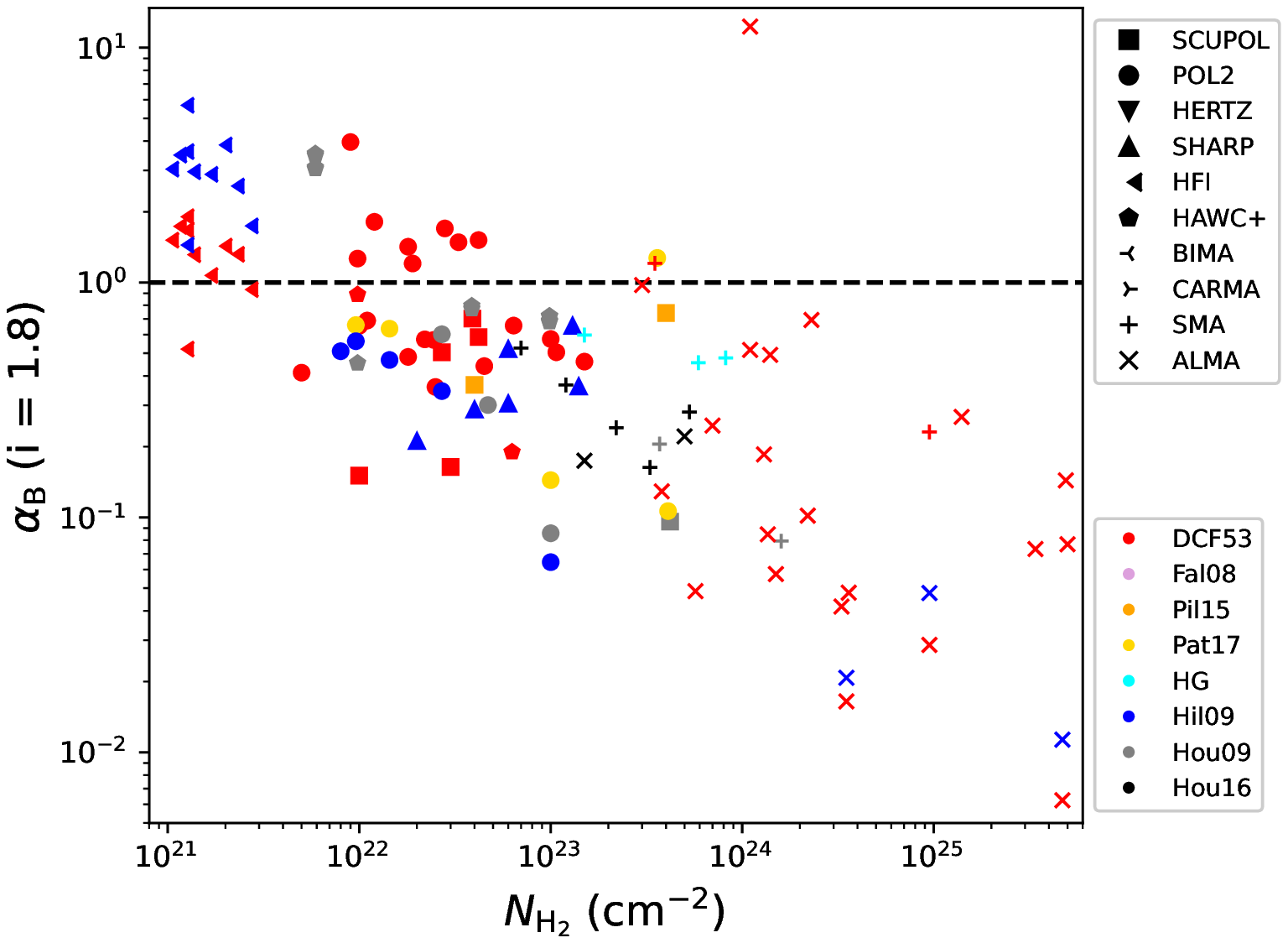}{0.45\textwidth}{(a)}
 \fig{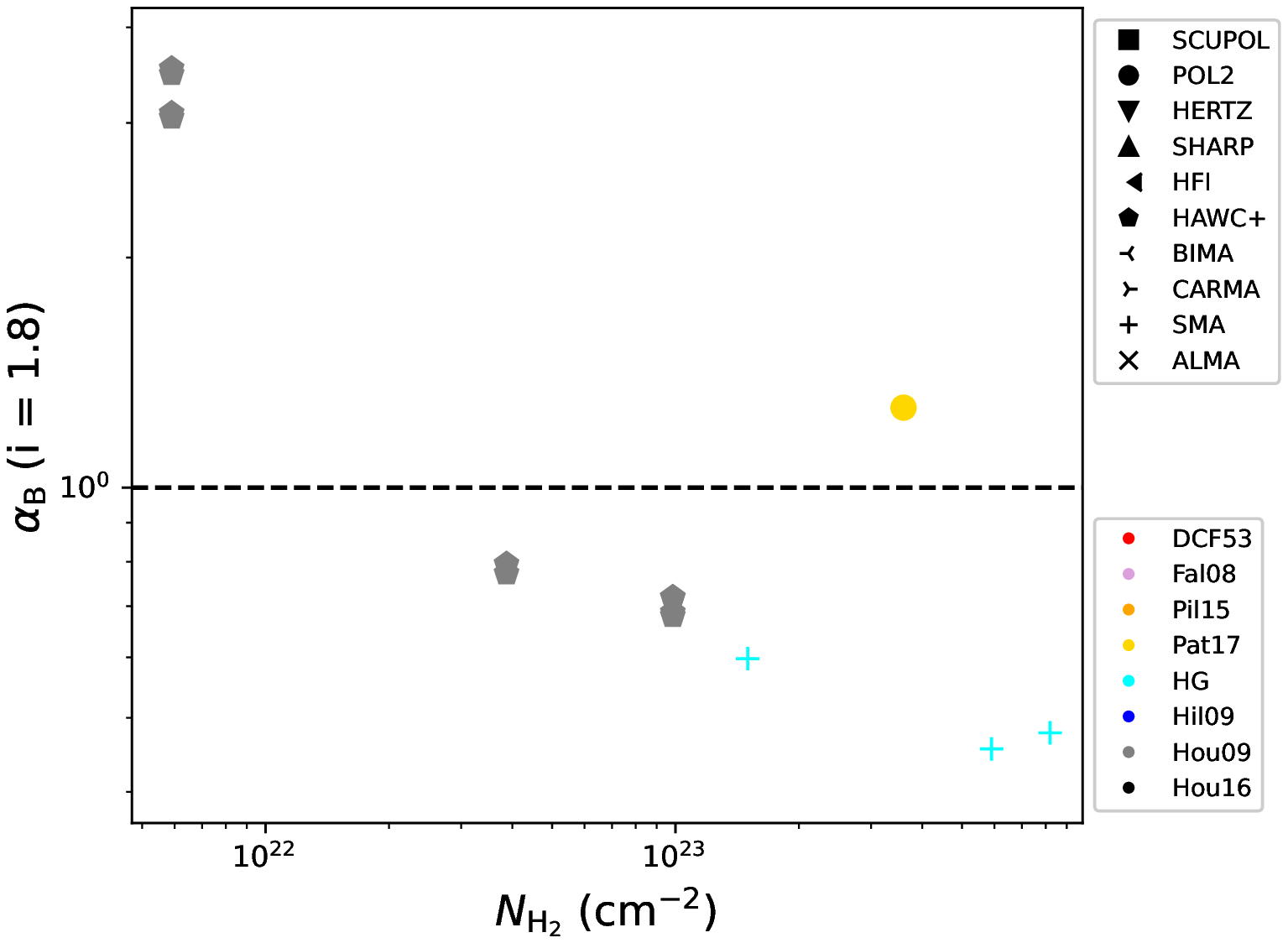}{0.45\textwidth}{(b)}
 }
\caption{The relation between magnetic virial parameter and column density for (a) a complete collection of the selected sample and (b) a subset of sample with hourglass field. Different symbols represent different instruments. Different colors represent different methods. \label{fig:alpha_N}}
\end{figure}

As mentioned before, the magnetic virial parameter can be reduced to the ratio between the magnetic field strength and the column density. Assuming $n \propto r^{-1.8}$ (see Section \ref{sec:nr}), we calculate the magnetic virial parameter with the estimated 3D total magnetic field strength and show its relation with the column density in Figure \ref{fig:alpha_N}(a). The transition from magnetically dominant at low column density to gravitationally dominant at high column density is also seen in the $\alpha_{\mathrm{B}} - N$ relation, where the magnetic virial parameter decreases (i.e., magnetic critical parameter increases) as a function of increasing column density. 

There could also be problems with the trend seen in Figures \ref{fig:B_N} and Figure \ref{fig:alpha_N}(a). In the estimation of the magnetic field strength, we adopted the correction factors for strong field cases that the turbulent magnetic energy and the turbulent kinetic energy are in approximate equipartition \citep{2021ApJ...919...79L}. However, if the strong field assumption is not satisfied, the magnetic field strength could be overestimated by the DCF method. Hourglass field morphology is a signpost for strong magnetic fields. In Figure \ref{fig:alpha_N}(b), we show the $\alpha_{\mathrm{B}} - N$ relation for a subset of samples with hourglass magnetic field morphology \citep[NGC1333 IRAS4A, NGC2024 FIR5, IRAS16293A, L1157, G240, L1448IRAS2, G31.41, OMC1, NGC6334I, ][]{2002ApJ...566..925L, 2006Sci...313..812G, 2007AJ....133.1012V, 2009ApJ...696..567H, 2009ApJ...706.1504H, 2009ApJ...707..921R, 2013ApJ...777..112P, 2013ApJ...769L..15S, 2014ApJ...794L..18Q, 2014ApJ...792..116Z, 2015Natur.520..518L, 2017ApJ...846..122P, 2019ApJ...872..187C, 2019ApJ...879...25K, 2019AandA...630A..54B, 2021ApJ...912..159P}. The decreasing $\alpha_{\mathrm{B}}$ with increasing $N$ is also seen in these hourglass field samples, which confirms the trend seen in Figures \ref{fig:B_N} and \ref{fig:alpha_N}(a). It should be noted that although hourglass field is a sign for strong magnetic fields, the lack of hourglass shape in other samples does not necessarily mean the field is weak. 
%However, if our samples deviate from the strong field assumption, the magnetic field strength could be overestimated and the transition between sub-critical and super-critical states could be at a lower column density.

Another uncertainty may come from the possible significant overestimation of angular dispersion at scales smaller than 0.1 pc due to line-of-sight signal averaging \citep{2021ApJ...919...79L}. In this case, the magnetic field would be overestimated and the gravity would be even more dominant in these small-scale and high-density regions. 

To conclude, we find the magnetic virial parameter decreases as the column density increases in the sample. This is in agreement with results of previous Zeeman studies \citep{2012ARAandA..50...29C} and studies of the relative orientation between magnetic fields and dense structures \citep{2021Galax...9...41L}. The dynamical states of the collected sample seem to be consistent with the prediction of the ambipolar diffusion process in strong field models \citep{1999ASIC..540..305M}, where the sub-critical clouds gradually lose magnetic flux via ambipolar diffusion, then become super-critical and collapse. Alternatively, the dissipation of magnetic flux at higher densities may also be explained by the process of magnetic reconnection \citep{1999ApJ...517..700L}. On the other hand, the mass flow along magnetic field lines increases the mass but not the magnetic flux in denser regions, which could also lead to the decreasing magnetic virial parameter at higher densities.
%We confirm that the dense structures shift from super-virial/sub-critical to sub-virial/super-critical as column density increases. 

\subsection{Comparing magnetic field with turbulence}

A major issue in studies of magnetic fields is to assess the relative importance of the turbulence and the magnetic field, which is usually done by comparing the magnetic energy with the turbulent kinetic energy or comparing the Alfv\'{e}n velocity with the turbulent velocity dispersion. 

As discussed in \citet{2021ApJ...919...79L}, because of the DCF assumption that the turbulent magnetic energy equals the turbulent kinetic energy, the total magnetic energy estimated from the DCF method would be always greater than the turbulent kinetic energy. Thus, we cannot properly compare the relative importance between the turbulence and the total magnetic field with estimations from the DCF method.

On the other hand, the comparison between the turbulence and the uniform magnetic field is done by estimating the 3D Alfv\'{e}nic Mach number
\begin{equation}
M_{\mathrm{A,3D}} = \frac{\sqrt{3}\delta v_{\mathrm{los}}\sqrt{\mu_0 \rho }}{B^{\mathrm{u}}_{3D}}.
\end{equation}
Because of the energy equipartition assumption of the DCF method, comparing the turbulence with the uniform magnetic field is equivalent to compare the turbulent and uniform components of the magnetic field. Therefore, measuring $M_{\mathrm{A,3D}}$ is equivalent to measuring the 3D turbulent-to-uniform magnetic field strength ratio $B^{\mathrm{t}}_{\mathrm{3D}}/B^{\mathrm{u}}_{3D}$. The relation between $B^{\mathrm{t}}_{\mathrm{3D}}/B^{\mathrm{u}}_{3D}$ and the referred turbulent-to-uniform magnetic field strength ratio  $B^{\mathrm{t}}_{\mathrm{pos\perp}}/B^{\mathrm{u}}_{\mathrm{pos}}$ in the DCF equation is 
\begin{equation}
B^{\mathrm{t}}_{\mathrm{pos\perp}}/B^{\mathrm{u}}_{\mathrm{pos}} = \frac{1}{\sqrt{3}} \frac{4}{\pi} B^{\mathrm{t}}_{\mathrm{3D}}/B^{\mathrm{u}}_{3D}
\end{equation}
where $B^{\mathrm{u}}_{\mathrm{3D}} = \frac{4}{\pi}B^{\mathrm{u}}_{\mathrm{pos}}$ and $B^{\mathrm{t}}_{\mathrm{3D}} = \sqrt{3}B^{\mathrm{t}}_{\mathrm{pos\perp}}$. A super-Alfv\'{e}nic state ($M_{\mathrm{Alf,3D}} > 1$, $B^{\mathrm{t}}_{\mathrm{3D}}/B^{\mathrm{u}}_{3D} > 1$, or $B^{\mathrm{t}}_{\mathrm{pos\perp}}/B^{\mathrm{u}}_{\mathrm{pos}} > 0.74$) indicates the turbulence is stronger than the uniform magnetic field, and vice versa. It should be noted that the energy equipartition assumption may not be satisfied in weak field cases \citep{2021ApJ...919...79L} and $B^{\mathrm{t}}_{\mathrm{3D}}/B^{\mathrm{u}}_{3D}$ could be smaller than $M_{\mathrm{A,3D}}$ in such cases. 

Another assumption of the DCF method is that the $B^{\mathrm{t}}_{\mathrm{pos\perp}}/B^{\mathrm{u}}_{\mathrm{pos}}$ is traced with the observed angular dispersion in forms of $\delta (\tan \phi)$ or $\delta \phi$ (in small angle approximation). However, the angular dispersion in the form of $\delta (\tan \phi)$ usually has large scatters and generally does not correctly estimate $B^{\mathrm{t}}_{\mathrm{pos\perp}}/B^{\mathrm{u}}_{\mathrm{pos}}$ \citep{2021ApJ...919...79L}. On the other hand, if the $B^{\mathrm{t}}_{\mathrm{pos\perp}}/B^{\mathrm{u}}_{\mathrm{pos}}$ is perfectly traced by $\delta \phi$, $B^{\mathrm{t}}_{\mathrm{pos\perp}}/B^{\mathrm{u}}_{\mathrm{pos}} \sim 0.74$ would correspond to $\delta \phi \sim 42.1\degr$. However, using $\delta \phi$ to trace $B^{\mathrm{t}}_{\mathrm{pos\perp}}/B^{\mathrm{u}}_{\mathrm{pos}}$ is only valid when $\delta \phi_{\mathrm{obs}} \lesssim 25\degr$ ($\sim$0.44) \citep{2021ApJ...919...79L}. When $\delta \phi_{\mathrm{obs}} > 25\degr$, the $B^{\mathrm{t}}_{\mathrm{pos\perp}}/B^{\mathrm{u}}_{\mathrm{pos}}$ can be significantly underestimated by $\delta \phi_{\mathrm{obs}}$. The correction factor between $\delta \phi$ and $B^{\mathrm{t}}_{\mathrm{pos\perp}}/B^{\mathrm{u}}_{\mathrm{pos}}$ is $\sim$0.25 for $\delta \phi_{\mathrm{obs}} \lesssim 25\degr$ \citep{2021ApJ...919...79L}. Thus, $B^{\mathrm{t}}_{\mathrm{pos\perp}}/B^{\mathrm{u}}_{\mathrm{pos}} \sim 0.74$ corresponds to $\delta \phi \sim 10.5\degr$ ($\sim$0.18). If the energy equipartition assumption is also satisfied, $\delta \phi > 10.5\degr$ would indicate a super-Alfv\'{e}nic state, and vice versa. Note that this is a statistical relation and may not be applicable to individual cases due to the projection effect and the scatter on the relation between the angular dispersion and the field strength ratio \citep{2021ApJ...919...79L}. 

The ADF method proposed to estimate the turbulent-to-ordered field strength ratio $(\langle B_{\mathrm{t}}^2 \rangle / \langle B_0^2\rangle)^{\frac{1}{2}}$ by fitting the ADFs. However, because the angular difference between two polarization angle estimates is fixed to be in the range of -90$\degr$ to 90$\degr$ \citep{2009ApJ...696..567H, 2009ApJ...706.1504H}, the upper limit of the derivable $(\langle B_{\mathrm{t}}^2 \rangle / \langle B_0^2\rangle)^{\frac{1}{2}}$ is 0.76 \citep{2021ApJ...919...79L}. There is no correction factors available between the $(\langle B_{\mathrm{t}}^2 \rangle / \langle B_0^2\rangle)^{\frac{1}{2}}$ derived from the ADF method and the $B^{\mathrm{t}}_{\mathrm{pos\perp}}/B^{\mathrm{u}}_{\mathrm{pos}}$, so it is unknown how accurately the ADF method estimates the turbulent-to-ordered field strength ratio. If the ratio between $(\langle B_{\mathrm{t}}^2 \rangle / \langle B_0^2\rangle)^{\frac{1}{2}}$ and $B^{\mathrm{t}}_{\mathrm{pos\perp}}/B^{\mathrm{u}}_{\mathrm{pos}}$ is greater than 1, the maximum derivable  $B^{\mathrm{t}}_{\mathrm{pos\perp}}/B^{\mathrm{u}}_{\mathrm{pos}}$ from the ADF method would be smaller than 0.74. In this case, the uniform magnetic energy estimated from the ADF method would also be always greater than the turbulent kinetic energy and the relative importance between the turbulence and the uniform magnetic field cannot be properly compared. 
%the ADF method does not correctly estimate $B^{\mathrm{t}}/B^{\mathrm{u}}$ when $B^{\mathrm{t}}/B^{\mathrm{u}} > 1$. To conclude,

\begin{figure}[!htbp]
 \gridline{\fig{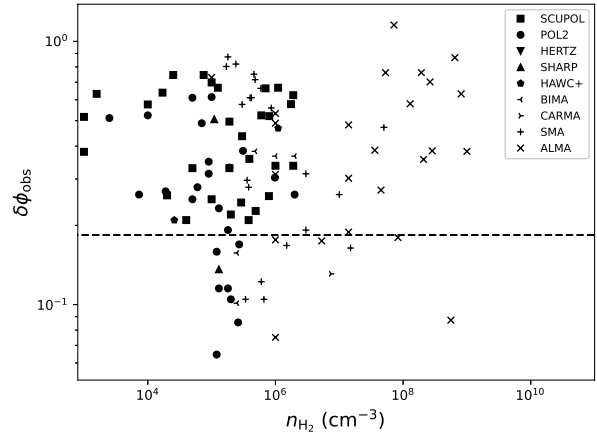}{0.45\textwidth}{}}
\caption{The relation between the observed angular dispersion and the volume density for the selected sample. Different symbols represent different instruments. The dashed line of $\delta \phi = 10.5\degr$ (0.18) indicates a trans-Alfv\'{e}nic state. \label{fig:ang_n}}
\end{figure}

Figure \ref{fig:ang_n} shows the relation between the observed angular dispersion and the volume density. We only plot the sources with $\delta \phi_{\mathrm{obs}}$ estimations. Data points from Planck observations are not plotted because no correction factors between $\delta \phi_{\mathrm{obs}}$ and $B^{\mathrm{t}}_{\mathrm{pos\perp}}/B^{\mathrm{u}}_{\mathrm{pos}}$ exist at the corresponding spatial scales. The $\delta \phi_{\mathrm{obs}}$ show large scatters. There is no strong relations between $\delta \phi_{\mathrm{obs}}$ and $n_{\mathrm{H_2}}$ in Figure \ref{fig:ang_n}. The low-density (e.g., $n_{\mathrm{H_2}} < 10^{5}$ cm$^{-3}$) structures and high-density (e.g., $n_{\mathrm{H_2}} > 10^{7}$ cm$^{-3}$) structures seem to be overall super-Alfv\'{e}nic, but it might be alternatively explained by a lack of samples at these density ranges. The average $\delta \phi_{\mathrm{obs}}$ is $\sim$23$\degr$ $\pm$13$\degr$ ($\sim$0.40$\pm$0.22), suggesting a slight super-Alfv\'{e}nic state with $M_{\mathrm{A,3D}} \sim 2.2 \pm 1.2$. Considering that the energy equipartition assumption may not be satisfied in weak field cases and that the angular dispersion may underestimate the turbulent-to-ordered field strength ratio when $\delta \phi_{\mathrm{obs}} > 25\degr$, the average state could be more super-Alfv\'{e}nic. It should be noted that the uniform magnetic field strength in \citet{2021ApJ...919...79L} is defined as the strength of the mean magnetic field vector. If a different definition is adopted (e.g., the contribution from the large-scale field structure is removed), the derived Alfv\'{e}nic states could be different. Assuming that the contribution of the 3D large-scale field structure to the 3D angular dispersion is similar to that of the 2D relation \citep[a factor of $\sim$2.5, ][]{2021ApJ...919...79L}, the average $M_{\mathrm{A,3D}}$ would be $\sim 0.9 \pm 0.5$, which is approximately trans-Alfv\'{e}nic.
%Because the correction factor between $\delta \phi_{\mathrm{obs}}$ and $B^{\mathrm{t}}_{\mathrm{pos\perp}}/B^{\mathrm{u}}_{\mathrm{pos}}$ in \citet{2021ApJ...919...79L} is for the directly measured angular dispersion,

%Due to the limitation of the DCF assumptions and limitations on the angular dispersions, we refrain from comparing the magnetic field with the turbulence. Further theoretical efforts

%discuss about the correction factor. 
%So the relative importance of the turbulence and the uniform magnetic field is also expected to be traced by angular dispersions. 

%\section{Discussion} \label{section:discussion}

%\subsection{$B - n$ relation}

%\subsection{$B - N$ relation: comparing magnetic field with gravity}

%shortness of DCF method. upper limit of angular dispersion.

\section{Conclusion} \label{section:conclusion}
We compile all the previous DCF estimations with polarized dust emission observations and re-calculate the magnetic field strength of the selected samples using the new DCF correction factors in \citet{2021ApJ...919...79L}. We investigate the relative importance of the magnetic field compared to the gravity based on the $B - n$ relation and the $B - N$ relation of the collected samples. We investigate the relative importance between the magnetic field and turbulence with the observed angular dispersions. The conclusions of this paper are as following:
\begin{enumerate}
    \item The power-law index of the $B - n$ relation is estimated to be $\sim$0.57, which is between the values predicted by strong and weak field models. However, the observed spatial $B - n$ relation may not be comparable to the predicted temporal $B - n$ relation of theoretical models. The derived power-law index also has large uncertainties due to large scatters in the observed $B - n$ relation, which further makes the comparison unreliable.
    \item The $B - N$ relation indicates that the samples are magnetically sub-critical (i.e., magnetically super-virial) at low column densities, and gradually transits to magnetically super-critical (i.e., magnetically sub-virial) at higher column densities. We find a clear trend of decreasing magnetic virial parameter (i.e., increasing mass-to-flux-ratio in units of critical value) with increasing column density. This suggests that the magnetic field is strong enough to balance gravity at low column density, but becomes insufficient to prevent gravitational collapse as the column density increases. This also indicates that the magnetic flux is dissipated at higher column densities due to the effects of ambipolar diffusion or magnetic reconnection, and the mass accumulation in denser regions may be by mass flows along magnetic field lines, which increases the mass but not the magnetic flux at higher densities.
    \item Both sub-Alfv\'{e}nic and super-Alfv\'{e}nic states are found in the selected sample. The average state is approximately trans-Alfv\'{e}nic.
    %\item \textbf{The cross-term magnetic field may not need to be considered in the commonly studied dense molecular structures where self-gravity is important.}
\end{enumerate}

\acknowledgments 
%We are indebted to the anonymous referee whose constructive comments improved the presentation and clarity of the paper. 
We thank the anonymous referee for constructive comments that improved the clarity of this paper. J.L. thanks Prof. Martin Houde for helpful discussions on the value range of the angular difference, Mr. James Beattie and Prof. Christoph Federrath for insightful discussions on the differences between the global and local magnetic field, and also Mr. Raphael Skalidis for helpful discussions on the cross-term magnetic energy although Mr. Skalidis has different opinions on the cross-term magnetic field. J.L. acknowledges the support from the EAO Fellowship Program under the umbrella of the East Asia Core Observatories Association. K.Q. is supported by National Key R\&D Program of China No. 2017YFA0402600. K.Q. acknowledges the support from National Natural Science Foundation of China (NSFC) through grants U1731237, 11590781, and 11629302. This research made use of Matplotlib, a Python 2D plotting library for Python \citep{2007CSE.....9...90H}.
\software{Matplotlib \citep{2007CSE.....9...90H}.}

%\newpage

\appendix

\section{Should we consider the cross-term magnetic energy?}\label{app:cross}
The DCF method assumes an equipartition between the turbulent magnetic energy and the turbulent kinetic energy. Recently, \citet{2021A&A...647A.186S} proposed that the cross-term (or coupling-term) magnetic energy dominates the turbulent magnetic energy in sub-Alfv\'{e}nic cases and thus the assumption of an equipartition between the cross-term magnetic energy and the turbulent kinetic energy should replace the DCF assumption. Here we discuss why the new energy equipartition assumption in \citet{2021A&A...647A.186S} may be inappropriate in the commonly studied star-forming molecular clouds/clumps/cores.

Let us firstly assume that the magnetic field vector of a basic fluid element $\boldsymbol{B_{\mathrm{ele}}}$ is composed of a mean (i.e., uniform) magnetic field component $\boldsymbol{B^{\mathrm{u}}}$ and a turbulent magnetic field component $\boldsymbol{B^{\mathrm{t}}_{\mathrm{ele}}}$, i.e., $\boldsymbol{B_{\mathrm{ele}}} = \boldsymbol{B^{\mathrm{u}}} + \boldsymbol{B^{\mathrm{t}}_{\mathrm{ele}}}$. The mean magnetic field vector $\boldsymbol{B^{\mathrm{u}}}$ for the considered group of fluid elements is defined as $\boldsymbol{B^{\mathrm{u}}} = \langle \boldsymbol{B_{\mathrm{ele}}} \rangle $. 

Globally, the average total (rms) magnetic energy of a group of fluid elements is proportional to 
\begin{equation}
    \langle \boldsymbol{B_{\mathrm{ele}}}^2 \rangle  = (\boldsymbol{B^{\mathrm{u}}})^2 + 2\langle \boldsymbol{B^{\mathrm{u}}} \cdot \boldsymbol{B^{\mathrm{t}}_{\mathrm{ele}}} \rangle + \langle(\boldsymbol{B^{\mathrm{t}}_{\mathrm{ele}}})^2 \rangle.
\end{equation}
The cross-term component $2\langle \boldsymbol{B^{\mathrm{u}}} \cdot \boldsymbol{B^{\mathrm{t}}_{\mathrm{ele}}} \rangle$ can be reduced to $2\langle B^{\mathrm{u}} B^{\mathrm{t}}_{\mathrm{\parallel,ele}} \rangle = 2B^{\mathrm{u}} \langle B^{\mathrm{t}}_{\mathrm{\parallel,ele}} \rangle$, where $\boldsymbol{B^{\mathrm{t}}_{\mathrm{\parallel,ele}}}$ is a component of the turbulent magnetic field vector of a fluid element that is parallel to $\boldsymbol{B^{\mathrm{u}}}$ and $B^{\mathrm{t}}_{\mathrm{\parallel,ele}}$ is the strength of $\boldsymbol{B^{\mathrm{t}}_{\mathrm{\parallel,ele}}}$. Because $\boldsymbol{B^{\mathrm{t}}_{\mathrm{\parallel,ele}}}$ could be along $\boldsymbol{B^{\mathrm{u}}}$ or at the opposite direction of $\boldsymbol{B^{\mathrm{u}}}$, $B^{\mathrm{t}}_{\mathrm{\parallel,ele}}$ could be either positive or negative. Along the $\boldsymbol{B^{\mathrm{u}}}$ direction, we have $B_{\mathrm{\parallel,ele}} = B^{\mathrm{u}} + B^{\mathrm{t}}_{\mathrm{\parallel,ele}}$ and $B^{\mathrm{u}} = \langle B_{\parallel,\mathrm{ele}} \rangle $, where $\boldsymbol{B_{\mathrm{\parallel,ele}}}$ is a component of the magnetic field vector of a fluid element that is parallel to $\boldsymbol{B^{\mathrm{u}}}$ and $B_{\mathrm{\parallel,ele}}$ is the strength of $\boldsymbol{B_{\mathrm{\parallel,ele}}}$. Then $\langle B^{\mathrm{t}}_{\mathrm{\parallel,ele}} \rangle = \langle B_{\mathrm{\parallel,ele}} \rangle - \langle B^{\mathrm{u}} \rangle = 0$. Thus, the cross-term component $2\langle \boldsymbol{B^{\mathrm{u}}} \cdot \boldsymbol{B^{\mathrm{t}}_{\mathrm{ele}}} \rangle$ for a group of fluid elements in the considered space would be 0 by definition, no matter whether the perturbed fluid is compressible or incompressible. It is clear that the cross-term magnetic energy is unimportant globally.
%So the total magnetic energy of a considered astronomical structure would equal the sum of the mean magnetic field energy and the turbulent magnetic field energy, where there is no room for a cross-term magnetic field energy. 
%e.g., Let us consider a sample 1D magnetic field array with elements 1, 1, 2, and 2 (with respect to unit strength). The mean field $B^{\mathrm{u}}$ is 1.5. The average total field strength is $\sim$1.58. The average turbulent field strength is 0.5. 
%= \langle ( \boldsymbol{B^{\mathrm{u}}} + \boldsymbol{B^{\mathrm{t}}_{\mathrm{ele}}})^2 \rangle

Locally, the total magnetic energy of a fluid element is proportional to 
\begin{equation}
    \boldsymbol{B_{\mathrm{ele}}}^2  = (\boldsymbol{B^{\mathrm{u}}})^2 + 2 \boldsymbol{B^{\mathrm{u}}} \cdot \boldsymbol{B^{\mathrm{t}}_{\mathrm{ele}}} + (\boldsymbol{B^{\mathrm{t}}_{\mathrm{ele}}})^2,
\end{equation}
where $2\boldsymbol{B^{\mathrm{u}}} \cdot \boldsymbol{B^{\mathrm{t}}_{\mathrm{ele}}} = 2B^{\mathrm{u}} B^{\mathrm{t}}_{\mathrm{\parallel,ele}}$ can be positive or negative. In strong field cases ($B^{\mathrm{t}}_{\mathrm{ele}} << B^{\mathrm{u}}$), the cross-term magnetic energy clearly dominates the turbulent magnetic energy. Thus, one can conclude that the cross-term magnetic energy matters a lot locally in strong field cases. 

How do we know whether the cross-term magnetic energy should be considered ``local'' or ``global'' in practice? The critical point might be determined by the turbulent correlation length \citep{2009ApJ...696..567H, 2009ApJ...706.1504H}. If the cross-term magnetic energy is averaged over scales larger than the turbulent correlation scale, we would expect that $B^{\mathrm{t}}_{\mathrm{\parallel,ele}}$ is random and the averaged cross-term magnetic energy is 0. On the other hand, if the averaging is done below the turbulent correlation scale, the value of $B^{\mathrm{t}}_{\mathrm{\parallel,ele}}$ within the turbulent cell would be likely correlated with each other and take the same sign \citep[i.e., positive or negative, ][]{2020MNRAS.498.1593B}. Thus, the averaged cross-term magnetic energy would not be 0 within the turbulent cell. 

Several numerical studies have looked into the issue of the cross-term magnetic energy \citep{2016JPlPh..82f5301F, 2020MNRAS.498.1593B, 2021arXiv210910925S}. The cross-term magnetic energy was found to be comparable to the turbulent kinetic energy in these works. However, the turbulence in these simulations is injected at scales comparable to the whole simulation box. Thus, the turbulent correlation scale would be comparable to the size of the simulation box \citep{2019MNRAS.488.2493B} and the cross-term magnetic energy should be considered ``local'' in these simulations. In this sense, the turbulence in these simulations would more resemble the interstellar supersonic turbulence in diffuse interstellar medium. At much smaller scales (e.g., within molecular clouds/clumps/cores), the ordered magnetic field of the particular structures is actually part of the turbulent magnetic field on larger scales \citep{2016JPlPh..82f5301F}. For the denser structures at these smaller scales, the subsonic cascaded turbulence \citep{2021NatAs...5..365F} or \textbf{the turbulence directly driven by stellar feedbacks of forming stars} should be considered instead of the large-scale supersonic turbulence. Observationally, the average turbulent correlation length within the molecular dense structures estimated with the ADF method is $\sim$0.04 pc (see Appendix \ref{app:dcfref}), which is smaller than the typical scale of molecular dense cores ($\sim$0.1 pc). Thus, the cross-term magnetic field in these commonly studied molecular star-forming clouds/clumps/cores is more likely ``global'' than ``local''. 
%

%simulations underestimate the turbulent magnetic energy because smaller than the turbulent correlation length.

Another problem of these simulations is that they are without gravity. In dense star-forming regions where self-gravity is significant, there could be strong ordered velocity field (e.g., infall and rotation) due to gravity. Even if the cross-term magnetic field should be considered in some cases, the cross-term (or coherent) velocity field should also be considered in the local energy budget for similar reasons. Thus, it may not be valid to simply equate the turbulent kinetic energy with the cross-term magnetic energy in regions where self-gravity is important. 

\section{Physical properties of the selected sample}\label{app:dcfref}
Here we list the basic physical properties, the referenced DCF estimations, and our new DCF estimations of the selected sample in Table \ref{tab:sources}. There are \textbf{230} measurements in total. As this paper is not intended to present a detailed review for each source, we do not check the consistency of source names and only adopt the name in the original reference. As we do not have the reduced polarization dataset of the selected sample for examination, we choose to trust the referenced papers and do not check whether there are sufficient polarization detections for a DCF analysis for each source \citep[i.e., size of polarization detection area greater than $\sim$2-4 times of the beam size, ][]{2021ApJ...919...79L}. The polarized dust emissions of the selected sample were observed with SCUPOL and POL-2 on JCMT \citep{2003MNRAS.340..353G, 2016SPIE.9914E..03F}, HERTZ and SHARP on CSO \citep{1998ApJ...504..588D, 2008ApOpt..47..422L}, HFI on Planck \citep{2010A&A...520A...9L}, HAWC+ on SOFIA \citep{2018JAI.....740008H}, BIMA, CARMA, SMA, and ALMA.

%%\colhead{$M_{\mathrm{A,3D}}$}  &
\begin{longrotatetable}
\movetabledown=0.5in
\begin{deluxetable}{ccccccccccccccccc}
\tablecaption{DCF sample \label{tab:sources}}
\tablecolumns{17}
\tablewidth{0pt}
\tabletypesize{\scriptsize}
\tablehead{
\colhead{Name} &
\colhead{Instrument} &
\colhead{Method}  &
\colhead{$r$}  &
\colhead{$M$}  &
\colhead{$n_{\mathrm{H_2}}$}  &
\colhead{$N_{\mathrm{H_2}}$}  &
\colhead{$\delta v_{\mathrm{los}}$}  &
\colhead{$\delta \phi_{\mathrm{obs}}$}  &
\colhead{$(\frac{\langle B_{\mathrm{t}}^2 \rangle}{\langle B_0^2\rangle})^{\frac{1}{2}}$ \tablenotemark{a}}  &
\colhead{$N_{\mathrm{adf}}$}  &
\colhead{$\delta_{\mathrm{adf}}$ \tablenotemark{b}} &
\colhead{$B^{\mathrm{u,ref}}_{\mathrm{pos}}$}  &
\colhead{$B^{\mathrm{u,est}}_{\mathrm{pos}}$}  &
\colhead{$B^{\mathrm{tot,est}}_{\mathrm{pos}}$}  &
\colhead{$\alpha_{\mathrm{B}}$}  &
\colhead{Reference}  \\
\colhead{}  &
\colhead{}  &
\colhead{}  &
\colhead{(pc)}  &
\colhead{($M_{\odot}$)}  &
\colhead{(cm$^{-3}$)}  &
\colhead{(cm$^{-2}$)}  &
\colhead{(km s$^{-1}$)}  &
\colhead{($\degr$)}  &
\colhead{}  &
\colhead{}  &
\colhead{(mpc)}  &
\colhead{($\mu$G)}  &
\colhead{($\mu$G)}  &
\colhead{($\mu$G)}  &
\colhead{}  &
\colhead{} 
}
\startdata
         SMM-NW &     SCUPOL & DCF53  &      ... &      ...     &   3.0 $\times$ 10$^{5}$ &      ...     &   1.10 &  25.0 &   ... &   ...&   ...&     1000 &      299 &      661 &   ... & \citet{ 2000ApJ...530L.115D} \\
           CB26 &     SCUPOL & DCF53  &      ... &      0.1     &   5.0 $\times$ 10$^{4}$ &      ...     &   0.14 &  18.9 &   ... &   ...&   ...&       74 &       21 &       45 &   ... & \citet{ 2001ApJ...561..871H} \\
           CB54 &     SCUPOL & DCF53  &      ... &     70.0     &   2.5 $\times$ 10$^{4}$ &      ...     &   0.38 &  42.7 &   ... &   ...&   ...&       60 &      ... &       39 &   ... & \citet{ 2001ApJ...561..871H} \\
           CG30 &     SCUPOL & DCF53  &      ... &      2.1     &   1.2 $\times$ 10$^{5}$ &      ...     &   0.14 &  38.2 &   ... &   ...&   ...&       16 &      ... &       36 &   ... & \citet{ 2001ApJ...561..871H} \\
 OrionB-NGC2071 &     SCUPOL & DCF53  &      ... &      ...     &   1.0 $\times$ 10$^{4}$ &      ...     &   0.42 &  33.0 &   ... &   ...&   ...&       56 &      ... &       35 &   ... & \citet{ 2002ApJ...569..304M} \\
    B335(CB199) &     SCUPOL & DCF53  &      ... &      5.0     &   1.9 $\times$ 10$^{6}$ &      ...     &   0.08 &  35.8 &   ... &   ...&   ...&      134 &      ... &       85 &   ... & \citet{ 2003ApJ...592..233W} \\
   CB230(L1177) &     SCUPOL & DCF53  &      ... &      7.0     &   8.0 $\times$ 10$^{5}$ &      ...     &   0.17 &  29.8 &   ... &   ...&   ...&      218 &      ... &      140 &   ... & \citet{ 2003ApJ...592..233W} \\
   CB244(L1262) &     SCUPOL & DCF53  &      ... &      1.5     &   1.8 $\times$ 10$^{6}$ &      ...     &   0.17 &  33.1 &   ... &   ...&   ...&      257 &      ... &      186 &   ... & \citet{ 2003ApJ...592..233W} \\
\enddata
\tablenotetext{a}{The turbulent-to-ordered magnetic field strength ratio derived from the ADF method.}
\tablenotetext{b}{The turbulent correlation length derived from the ADF method.}
\end{deluxetable}
\end{longrotatetable}

\end{CJK*}
\end{document}